\documentclass[aps,preprint,prd,showpacs,nofootinbib]{revtex4}

\usepackage{latexsym}
\usepackage{amsmath}
\usepackage{graphicx}
\usepackage{subfigure}
\usepackage{dcolumn}
\usepackage{bm}
\usepackage{amssymb}
\usepackage{color}
\usepackage{float}
\usepackage[colorlinks,linkcolor=magenta,anchorcolor=cyan,citecolor=blue]{hyperref}

\def\be{\begin{equation}}
\def\ee{\end{equation}}
\def\ba{\begin{aligned}}
\def\ea{\end{aligned}}

\def\lf{\left}
\def\rt{\right}

\begin{document}

\title{On primordial universe in anti-de Sitter landscape }

\author{Pu-Xin Lin$^{1,2}$\footnote{linpiaoxin17@mails.ucas.ac.cn}}
\author{Hai-Long Huang$^{1,3}$\footnote{huanghailong18@mails.ucas.ac.cn}}
\author{Jun Zhang$^{4}$\footnote{zhangjun@ucas.ac.cn}}
\author{Yun-Song Piao$^{1,3,4,5}$\footnote{yspiao@ucas.ac.cn}}

\affiliation{$^1$ School of Fundamental Physics and Mathematical
    Sciences, Hangzhou Institute for Advanced Study, UCAS, Hangzhou
    310024, China}

\affiliation{$^2$ Department of Physics, University of
Wisconsin-Madison, 1150 University Ave, Madison, WI 53706, U.S.A.}

\affiliation{$^3$ School of Physical Sciences, University of
Chinese Academy of Sciences, Beijing 100049, China}

\affiliation{$^4$ International Center for Theoretical Physics
    Asia-Pacific, Beijing/Hangzhou, China}

\affiliation{$^5$ Institute of Theoretical Physics, Chinese
    Academy of Sciences, P.O. Box 2735, Beijing 100190, China}

\begin{abstract}

How the spacetime evolved non-perturbatively in a landscape with
multiple anti-de Sitter (AdS) vacua, which is theoretically
well-motivated, has always been a matter of concern. As a step
towards this issue, we perform (3+1)D numerical relativity
simulations for the inhomogeneous universe in an AdS landscape,
and find that large inhomogeneity of scalar field can develop into
not only sphere-like bubbles, but also novel tube-like structures.
It is observed that the bubble or tube wall (across the potential
barrier) likely inflates as a quasi-dS space, while the different
regions separated by the walls are in different AdS vacua and will
collapse towards singularity.

%It is confirmed that in such AdS landscapes, cosmic inflation will
%inevitably appear.

\end{abstract}

\maketitle

\section{Introduction}

The success of inflation
\cite{Guth:1980zm,Linde:1981mu,Albrecht:1982wi,Starobinsky:1980te,Linde:1983gd},
consistent with current cosmological observations, suggests the
existence of a de Sitter (dS) phase in the very early stage of our
observable Universe. In string theory,
although it is possible that dS vacua exist
\cite{Kachru:2003aw,Kallosh:2019axr}, the construction of such
vacua in string landscape is not straightforward. There are
however valid concerns about the validity of such constructions in
string theory, notably the swampland conjecture
\cite{Ooguri:2006in,Obied:2018sgi}, see also
\cite{Brennan:2017rbf,Palti:2022edh} for recent reviews. In
contrast, anti-de Sitter (AdS) vacua are easy to construct and are
ubiquitous. Recently, it has been showed that AdS spacetime not
only plays crucial roles as insights into our Universe,
e.g.\cite{Chen:2020tes,Hartman:2020khs,Levine:2022wos,Cooper:2018cmb,Antonini:2022blk,Antonini:2022xzo}
but also have potential observable imprints
\cite{Ye:2020btb,Ye:2020oix,Jiang:2021bab,Ye:2021iwa}.

The initial state of the universe might be highly inhomogeneous
\cite{Feldbrugge:2017fcc}, i.e. the scalar field or spacetime
metric can exhibit non-perturbative inhomogeneities before a
region of space arrives at a certain vacuum, so that initially
well-defined background in such highly inhomogeneous states does
not exist. It is also often speculated that due to the past
incompleteness of inflation \cite{Borde:2001nh}, large
fluctuations of spacetime will be inevitably excited near the
initial cosmological singularity, in which case the initial
conditions of scalar fields and spacetime are chaotic
\cite{Linde:1983gd}. In such a non-perturbatively and highly
inhomogeneous spacetime, perturbative calculation is not
applicable any more.

In the past decades, numerical relativity (NR) has developed
significantly and has been applied to studies on the
merging of black hole binaries
\cite{Pretorius:2005gq,Campanelli:2005dd,Baker:2005vv}
%, see
%\cite{Lehner:2014asa,Cardoso:2014uka,Palenzuela:2020tga} for
%recent reviews,
and non-perturbative cosmologies,
e.g.\cite{Giblin:2015vwq,Bentivegna:2015flc,Musoke:2019ima,Macpherson:2016ict,Macpherson:2018btl}.
As pointed out, the study of AdS landscape is theoretically
well-motivated. It is well-known that the region in a single AdS
vacuum is unstable and will eventually collapse into a singularity
\cite{Abbott:1985kr}, see also e.g.\cite{Bizon:2011gg}.
Consequently, the non-perturbative evolution of the entire
spacetime in such landscape has continued to be a matter of
concern, e.g.\cite{Blanco-Pillado:2019tdf}.

%The collapse of AdS bubble has been studied earlier in
%Ref.\cite{Abbott:1985kr}. The nonlinear evolution of a weakly
%perturbed AdS spacetime has been also solved numerically in
%Refs.\cite{Bizon:2011gg,Jalmuzna:2011qw,Dias:2011ss,Buchel:2012uh},
%which, although are still limited to the perturbation-like
%studies, have found that such a AdS region is unstable under
%perturbations and will eventually collapse. As commented, see also
%\cite{Ooguri:2006in,Obied:2018sgi}, the string landscape
%inevitably contains lots of AdS vacua, however, in such a
%landscape, the evolution of the spacetime is far more complicated,
%so that the perturbative method is not sufficient to exhibit real
%and rich phenomena.

%However,
%in an AdS landscape, the dynamics and evolution of spacetime is
%far more complicated, exhibiting rich phenomena that has not been
%well explored thus far.

\begin{figure}
    \centering
    \includegraphics[width=16cm]{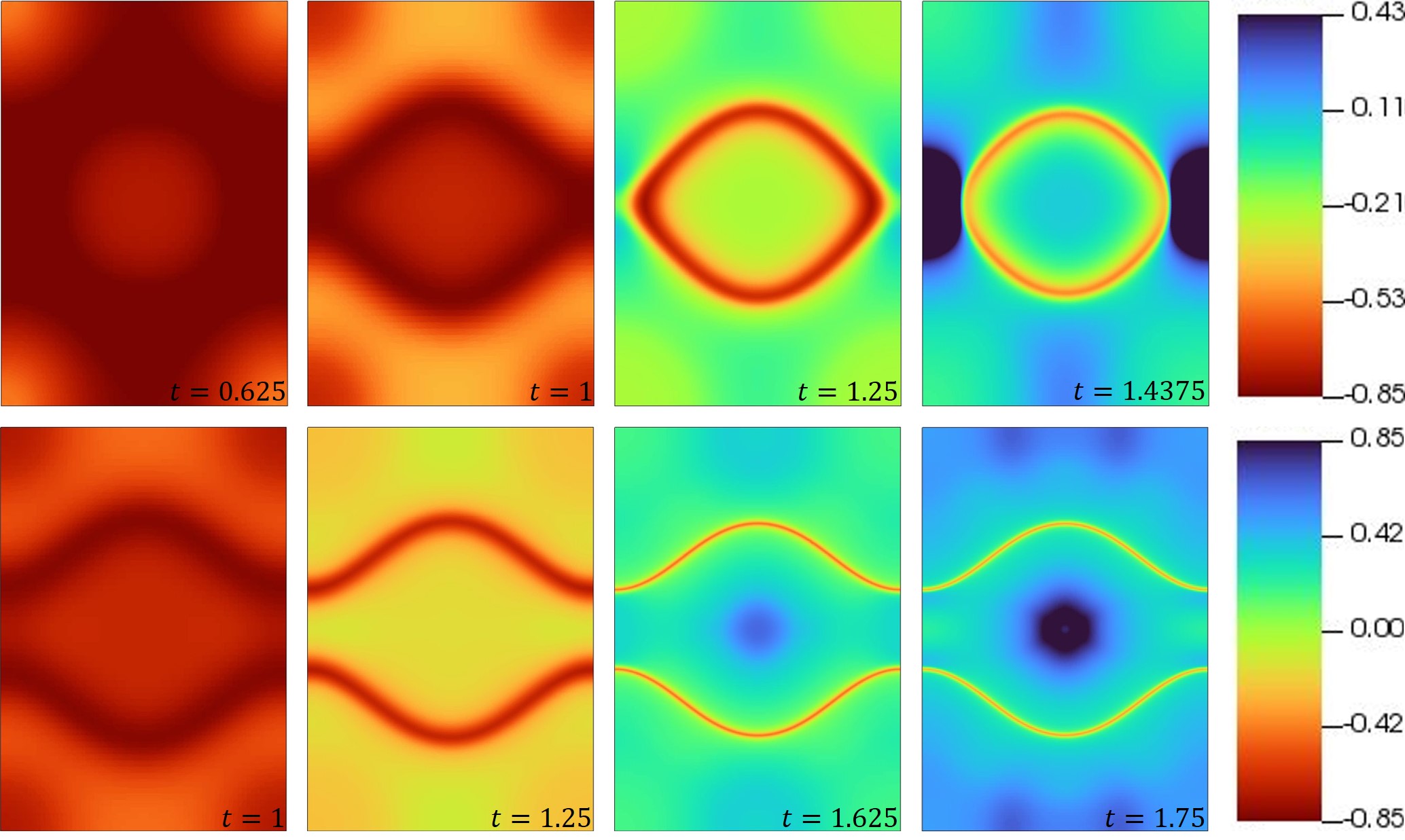}
\caption{ \textbf{Snapshots of spacetime.} The upper and lower
panel show how the sphere-like bubbles ($\phi_0=0.89$) and 
tube-like structures ($\phi_0=0.92$) form
respectively, which are seen in light of the plane passing the
center of the grid with normal $\vec{n}=(1,-1,0)$. Initially, the
whole space is expanding ($K<0$ at the red end of the colorbar).
After a period of evolution, different regions of space will be
separated by expanding walls (the bubble wall or tube-like
``ribbon" at which $K<0$) into different AdS vacua, while the
corresponding AdS regions will collapse ($K>0$ with a blue shifted
color).}
    \label{K_2d}
\end{figure}

As a step towards this issue, we perform (3+1)D NR simulations of
the ``universe" in a simplified AdS landscape. Despite the
application of NR simulations in cosmologies
\cite{Giblin:2015vwq,Bentivegna:2015flc,East:2015ggf,Clough:2016ymm,Lin:2021ubu},
few focused on the AdS landscape.
Specially, in Ref.~\cite{Lin:2021ubu}, we investigate the evolution of scalar field in an initial inhomogeneous
expanding Universe and how a patch of spacetime evolves into the dS vacua that should
happen at the early epoch of the Universe. In this letter, we further investigate the spacetime evolves in a landscape with multiple AdS vacua, which are more generic in string
theory. We explore potential new phenomenology and investigate whether a quasi-de Sitter space can emerge on the bubble wall separating different AdS regions.
%To our knowledge, this work is
%the first full relativistic simulation showing the
%non-perturbative evolution of spacetime in the AdS landscape.
Our simulation results show novel phenomena not captured by
conventional perturbative analyses, see Fig.\ref{K_2d}, which in
certain sense highlights the important role of NR in comprehending
the non-perturbative phenomena of spacetime.

\section{Method}

\begin{figure}
    \centering
    \includegraphics[width=10cm]{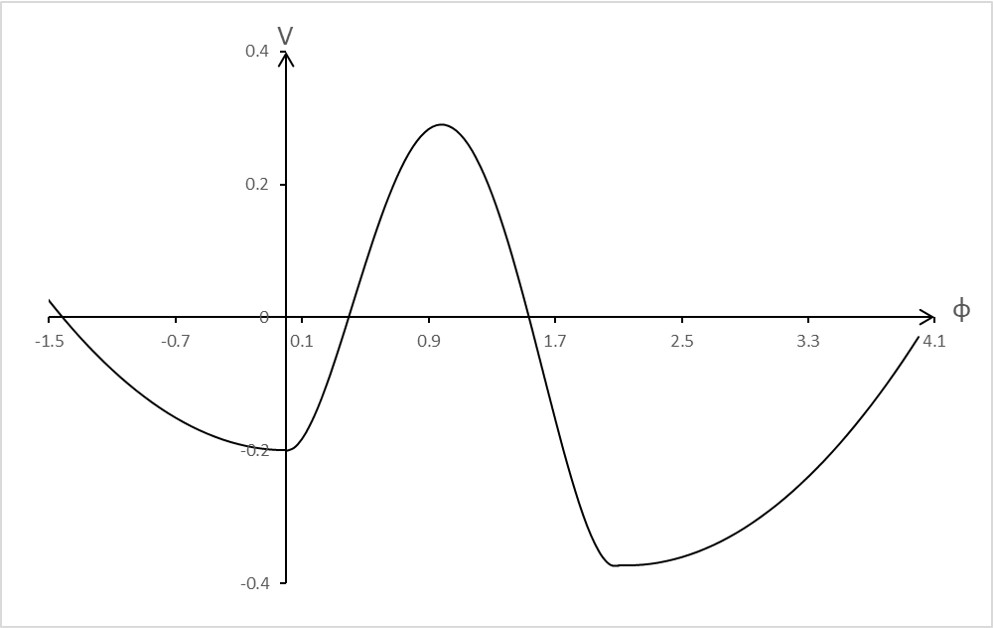}
\caption{\textbf{A simplified AdS landscape.} The effective
potential $V(\phi)$ has two AdS minima separated by a potential
barrier with positive energy. }
    \label{potential}
\end{figure}

Throughout this work, we consider a 1D potential as a
phenomenological simplification of the actual complex AdS
landscape, setting a fourth-order polynomial barrier between the
two AdS minima that have quadratic forms $\sim \phi^2$ (see
Fig.\ref{potential}), \be V\lf(\phi \rt) = \left\{\begin{aligned}
    &\frac{1}{2}m_1^2 { \lf( \phi - \phi_1 \rt) }^2 +V_1, &\phi<\phi_1,\\
   &\lambda\lf(\phi^2+{1\over \lambda}\rt)^2-2 \phi^3-{1\over \lambda}+V_1, \quad &\quad \phi_1\leq\phi\leq\phi_2, \\
    &\frac{1}{2}m_2^2 { \lf( \phi - \phi_2 \rt) }^2+V_2,&\phi_2<\phi,
\end{aligned}\right.
\label{V}\ee where $V_{1},V_2<0$. Here, we set $m_1^2=m_2^2=0.2$,
$\lambda=0.49$ ($c=\hbar=8\pi G=1$). We thus have the AdS-like vacua located at
$\phi_1=0$ and $\phi_2\simeq 2$, where the potential is
constructed to be differentiable, and the maximal value of
potential barrier is at $\phi_b\simeq 1$.

The large non-perturbative inhomogeneity of fields will be excited
inevitably near the initial singularity or the Planck scale,
giving rise to the possibility of initially configuration
$\partial_i\phi\partial^i\phi\lesssim 1$, or $\Delta\phi\lesssim
1$. As example, we set the initial inhomogeneity of scalar field
as \be \lf.\phi \rt|_{t=0} = \phi_0 + \Delta \phi
\sum_{\vec{x}=x,y,z}\cos \lf( \frac{2\pi \vec{x}}{L} \rt),
\label{phi}\ee where the amplitude of inhomogeneity is
$\Delta\phi= {\cal O}(0.1)$, the wavelength of inhomogeneity is
taken to be equal to the length of the periodic cubic region (in
our simulation $L=4$). In our simulation, we set the periodic
boundary condition\footnote{However, we also consider the
reflective boundary condition with the GRChombo code, and still
observed similar results. The details will be provided upon
request.}, and $\Delta\phi=0.4$ is the default value unless 
otherwise specified, with $\phi_0$ varying from 0.8 to 0.98.

We modify the BSSN based \cite{Baumgarte:1998te,Shibata:1995we} NR
package
GRChombo\footnote{http://www.grchombo.org\\https://github.com/GRChombo}
\cite{Clough:2015sqa} to perform the simulation. The NR
simulation is usually performed in the 3+1 decomposition context
for spacetime, with the metric as $g_{00} = -{\alpha}^2+{\beta}_i
{\beta}^i$, $g_{0i} = {\beta}_i$ and $g_{ij} = {\gamma}_{ij}$,
where $\alpha$ is the lapse parameter, ${\beta}^i$ the shift
vector and ${\gamma}_{ij}$ the spatial metric, while for the
numerical stability, the spatial metric $\gamma_{ij}$ must be
further factorized into its determinant $\chi$ (conformal factor)
and a comformal metric $\widetilde{\gamma}_{ij}=\chi
\gamma_{ij}$. The evolutions of $\widetilde{\gamma}_{ij}$, the
connections $\widetilde{\Gamma}^i= \widetilde{\gamma}^{jk}
\widetilde{\Gamma}^i_{jk}$ and the extrinsic curvature
$K_{ij}=\frac{1}{3}K\delta_{ij}+A_{ij}$ comply with the BSSN
equations \cite{Baumgarte:1998te,Shibata:1995we}.

Initially, the values of the BSSN variables are set as
$\widetilde{\gamma}_{ij}=\delta_{ij}$, $\widetilde{A}_{ij}=\chi
A_{ij}=0$, and $\chi=1$. In GRChombo code
\cite{Clough:2015sqa}, to satisfy the Hamiltonian constraint
involving the scalar field, we enforce the constraint by relaxing
the conformal factor in the light of ${\partial}_t \chi
=\mathcal{H}$ \footnote{This parabolic equation evolves $\chi$
into a steady state ${\partial}_t \chi =\mathcal{H}$. 
It enforces
the Hamiltonian constraint and thus prepares a proper initial
condition.
Here, 
the Hamiltonian constraint violation $\mathcal{H}$ is given by
\begin{equation}
{\cal H}=\tilde{D}^2\chi-\frac{5}{4\chi}\tilde{\gamma}^{ij}\tilde{D}_i\chi\tilde{D}_j\chi+\frac
{\chi}{2}\tilde{R}+\frac{1}{3}K^2-\frac{\chi^3}{2}\tilde{A}^{ij}
\tilde{A}_{ij}-\rho. 
\end{equation}
}, e.g. see recent
Refs.\cite{Clough:2016ymm,Lin:2021ubu}. In addition, we set the
initial scalar field at rest $\dot{\phi}=0$, satisfying the
momentum constraint and the initial expansion rate to be uniform
$K_{init} = -\sqrt{3\langle\rho\rangle}\approx
-1.08$, where $\langle\rho\rangle$ indicates the average 
initial erergy
density. 
% The local Hubble rate (reflecting local
% expanding rate of spacetime) is defined as
% \cite{East:2015ggf,Clough:2016ymm,Lin:2021ubu} \be
% H_{local}=\lim\limits_{V\rightarrow 0}\left({1\over
% V}\int{\rho(\phi)\over 3} dV\right)^{1/2}=-{K\over 3}.
% \label{local_H} \ee
In the special case of an isoptopic and homogeneous Universe, extrinsic curvature $K$ is related to the Hubble constant as $K=-3H$ with geodesic observers.

%In the simulation, since the evolutions of different local
%regions are different, it might be possible that some local
%regions would develop non-physical diverging metric components. It
%is thus important to ensure that our numerical calculation is
%still reliable. Here, it equals to require the Hamiltonian
%constraint conserved, i.e. $\mathcal{H}$ remains sufficiently
%small in accordance with the simulation precision for the results
%that we presented.

\section{Results}

\begin{figure}
    \centering
\includegraphics[width=14cm]{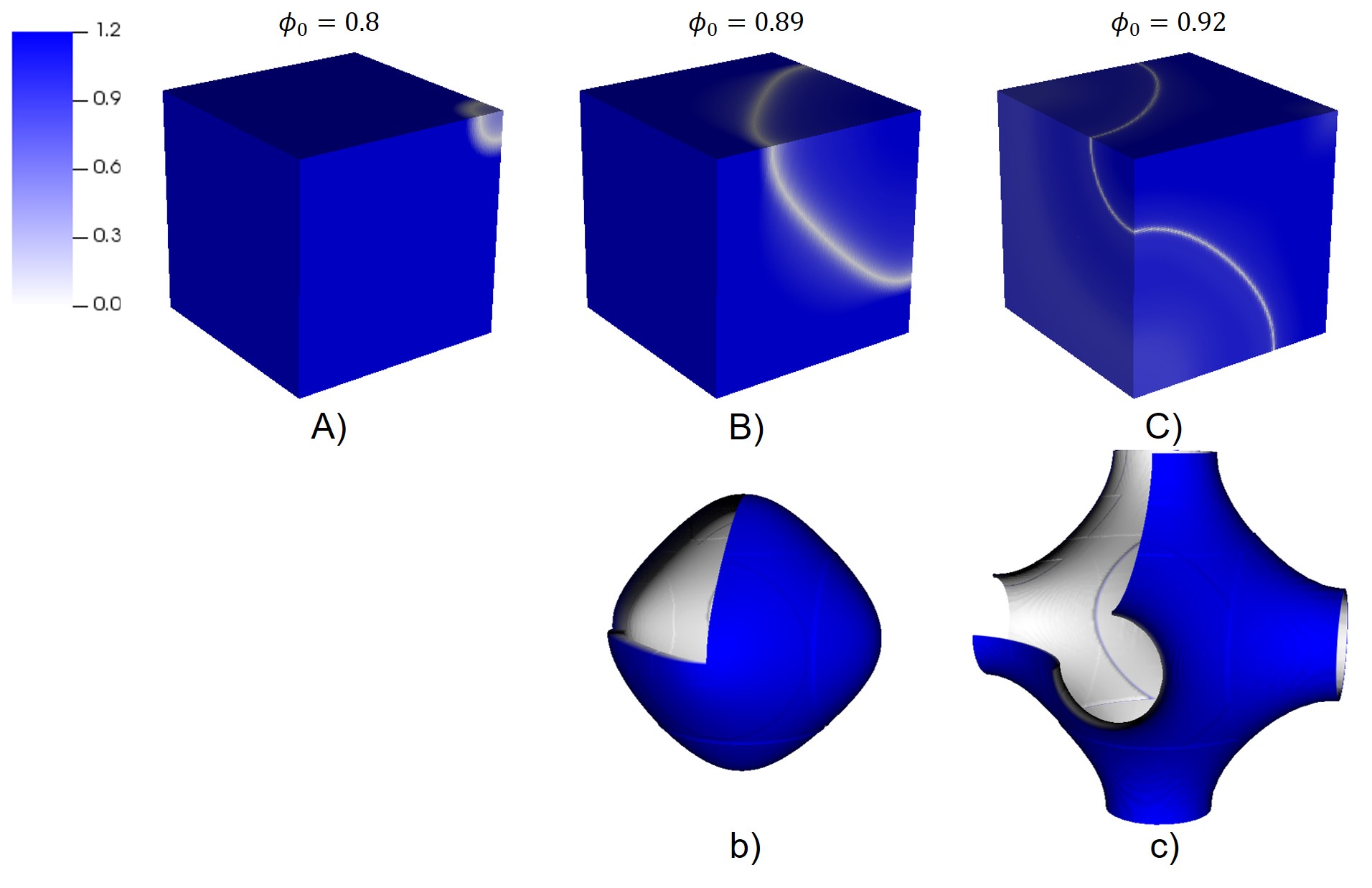}
\caption{\textbf{The value of $|\phi-\phi_b|$ (1/8 of the whole
computational grid).} In subfig-A, B and C, the lower left and
upper right regions correspond to the left and right side of the
potential barrier in Fig.\ref{potential}, respectively, and the
boundaries colored white mark the position of the barrier. In
subfig-A, the snapshot is taken shortly before the upper right AdS
region disappears. In subfig-B or C, the snapshot is taken when
the bubble or tube wall eventually arrives at rest. The subfig-b
and c show the interior of walls with 1/8 of the computational
grid removed. }
    \label{phi_cases}
\end{figure}

The simulation results are presented in Fig.\ref{K_2d} and
Fig.\ref{phi_cases}. The large inhomogeneity will spawn AdS
bubbles in a background of a different AdS vacuum. As expected, if
a bubble has subcritical initial radius, it will rapidly shrink
and collapse into a black hole\footnote{Using the Apparent Horizon Finder in GRChombo
code, we find that the mass of the black hole is $M\approx 5.7$, which is
associated with the apparent horizon and less than but converges
to the actual mass of black hole.}, and eventually the whole space
will evolve into a single AdS state and collapse integrally, see
Fig.\ref{phi_cases}-A. However, if the initial radii of the
bubbles are supercritical, the regions with different AdS vacua
will coexist, separated by not only near-spherical bubble walls
(Fig.\ref{phi_cases}-B) but also, interestingly, infinitely
extending tube-like walls (Fig.\ref{phi_cases}-C). Here, the
critical radius of bubble refers to the minimal radius of bubble
when the bubble phase can exist.

These tube-like walls can be understood more clearly in
Fig.\ref{K_2d}. In our simulation, we replace an infinite space
with the periodic boundary condition. This choice of boundary
condition is motivated as follows: inhomogeneities occur at
various scales, when considering evolution of those at a specific
length scale, it is natural to expect the extent of the modes goes
beyond just a single period.
%i.e., the periodic boundary condition
%is not really imposed but rather a consequence of restricting the
%scope of investigation to a region that has the same length scale
%as the inhomogeneity mode.
In light of this argument, the
``bubble" in the lower panel of Fig.\ref{K_2d} actually correspond
to the tube that extend along $\overrightarrow{x}(=x,y,z)$
directions through the region where the periodicity of the initial
perturbation modes persists. Such tube-like structure looks ``as
if" the adjacent spherical cells merged partly.
It is intriguing to consider the more generic setup with perturbations at numerous
wavelengths, a task beyond the scope of this letter, and to be pursed in future
work.

\begin{figure}
    \centering
\includegraphics[width=14cm]{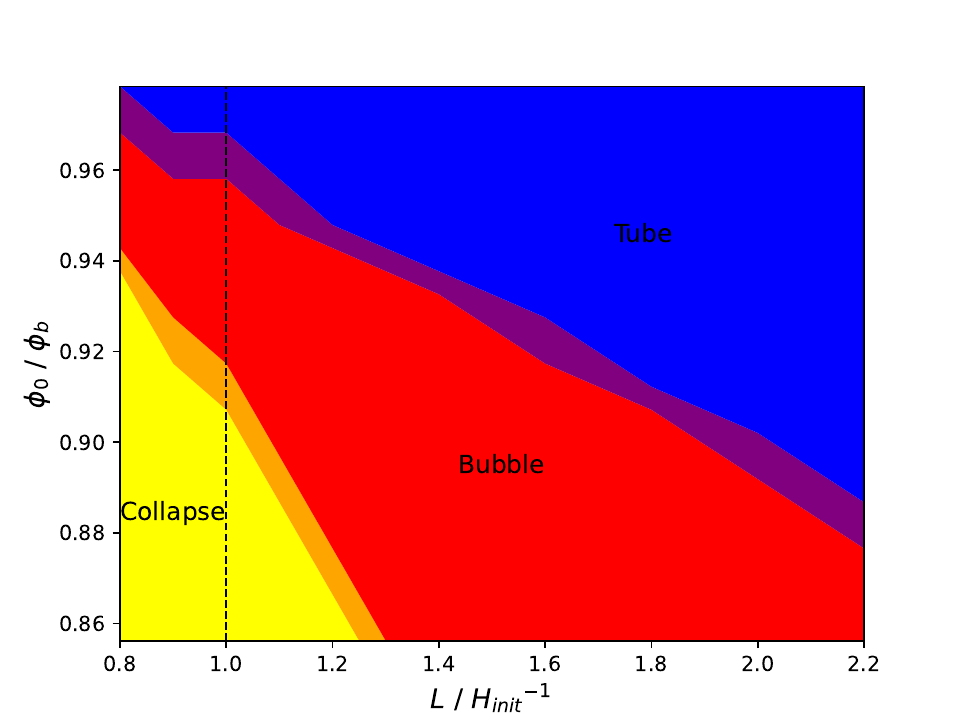}
\caption{\textbf{The phase diagram of bubble-tube Phase
Transition}. ``Collapse" represents the phase in which the whole
space will collapse integrally, ``Bubble" and ``Tube" represent
phases with existence of bubbles and tube-like structures
respectively. The boundaries between them (indicated by the orange
and purple strips) are not sharp since we cannot know which will
be the final state in our simulation precision. Here, we
mainly focus on the parameter space in which the wavelengths of
initial inhomogeneity are very close to the initial horizon,
$L\thickapprox H_{init}^{-1}$, in order to see the Phase
Transition between different phases.}
    \label{delta04}
\end{figure}

It is significant to check the effect of the wavelength $L$ of
inhomogeneity (equivalently $L/H_{init}^{-1}$). A phase diagram is
presented in Fig.\ref{delta04} \footnote{Here, we fix the
amplitude $\Delta\phi=0.4$ of initial inhomogeneity. However, we
actually also consider different $\Delta\phi=0.3, 0.5$, but we
find that the corresponding phase diagrams are essentially similar
to Fig.4, only the slope of boundary lines separating the
``Bubble" and ``Tube" phases are different.}. According to
Fig.\ref{delta04}, for $\phi_0/\phi_b\lesssim 0.9$, if
$L/H_{init}^{-1}\lesssim 1$ (i.e. the wavelength $L$ is
``subhorizonal"), eventually the whole space will collapse
integrally, while the bubble or tube phase comes into being only
when the initial inhomogeneity is ``superhorizonal" (the radius of
the bubble is supercritical and superhorizonal). It is also worth
noting that if $\phi_0/\phi_b\gtrsim 0.9$, i.e. $\phi_0$ is close
to the maximal value of potential barrier, we will see the bubble
or tube phase for $L/H_{init}^{-1}\lesssim 1$ (the radius of the
bubble is supercritical but subhorizonal).

Physically, our results can be qualitatively understood as
follows. Here, if a bubble has subcritical initial radius, it will
rapidly shrink and disappear, and eventually the whole space will
evolve into a single AdS state and collapse integrally, while if a
bubble has supercritical and superhorizonal initial radius, the
collapse of bubble wall will be prevented by the expansion of wall
spacetime, as expected theoretically in
Ref.\cite{Blanco-Pillado:2019tdf}, so a bubble phase will come
into being. As the initial radii of bubbles gets bigger (than a
certain critical value), the bubbles will get closer so that they
can partly join together through the tube naturally.

\begin{figure}[H]
    \centering
\includegraphics[width=14cm]{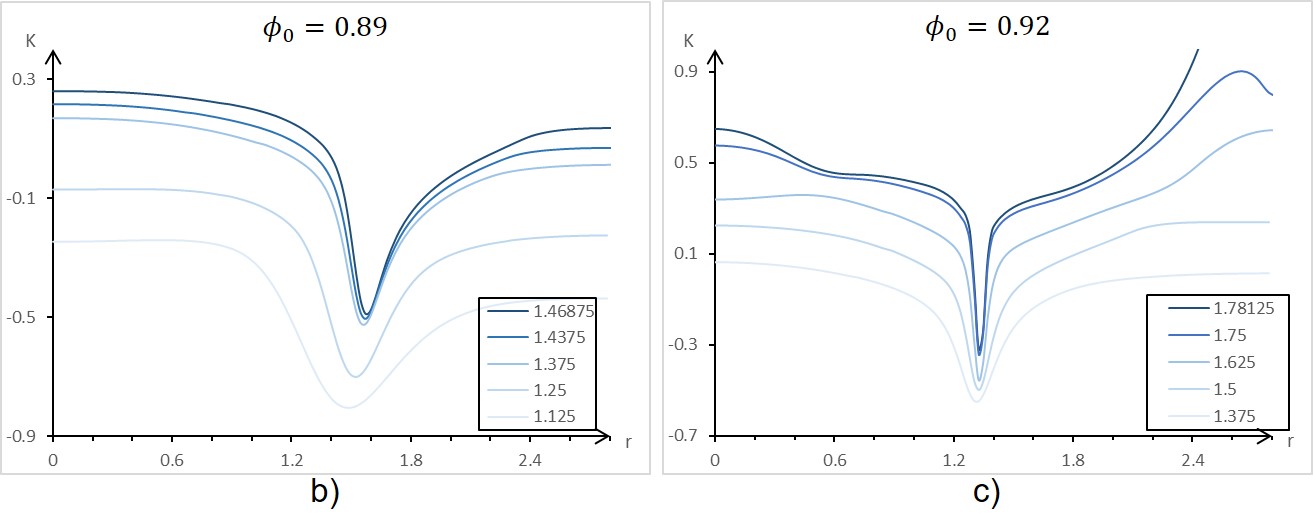}
\caption{\textbf{The evolution of extrinsic curvature scalar $K$.}
The horizonal axis is the line from cubic center to the middle
point on the edge of whole computational grids (upper right and
left vertexes of the boxes in Fig.\ref{phi_cases}). In subfig-b
and c, the bubble and tube walls form, and eventually the walls
will arrive a quasi-dS stationary profile ($K=const.<0$ on wall).}
    \label{K_line}
\end{figure}

Here, the initial ``universe" is set to be expanding homogeneously
(all $H_{local}=const.>0$), but eventually different regions will
tend to evolve into different AdS vacua and be separated by bubble
or tube walls. According to Fig.\ref{K_line}, we see that though
the AdS regions at both sides of the wall are collapsing, the
existing walls will still expand and eventually arrive at a
quasi-dS static profile, implying that the inflationary stage
responsible for our observable Universe might happen at such
walls, e.g.\cite{Vilenkin:1994pv,Linde:1994hy}.

\begin{figure}[H]
    \centering
\includegraphics[width=16cm]{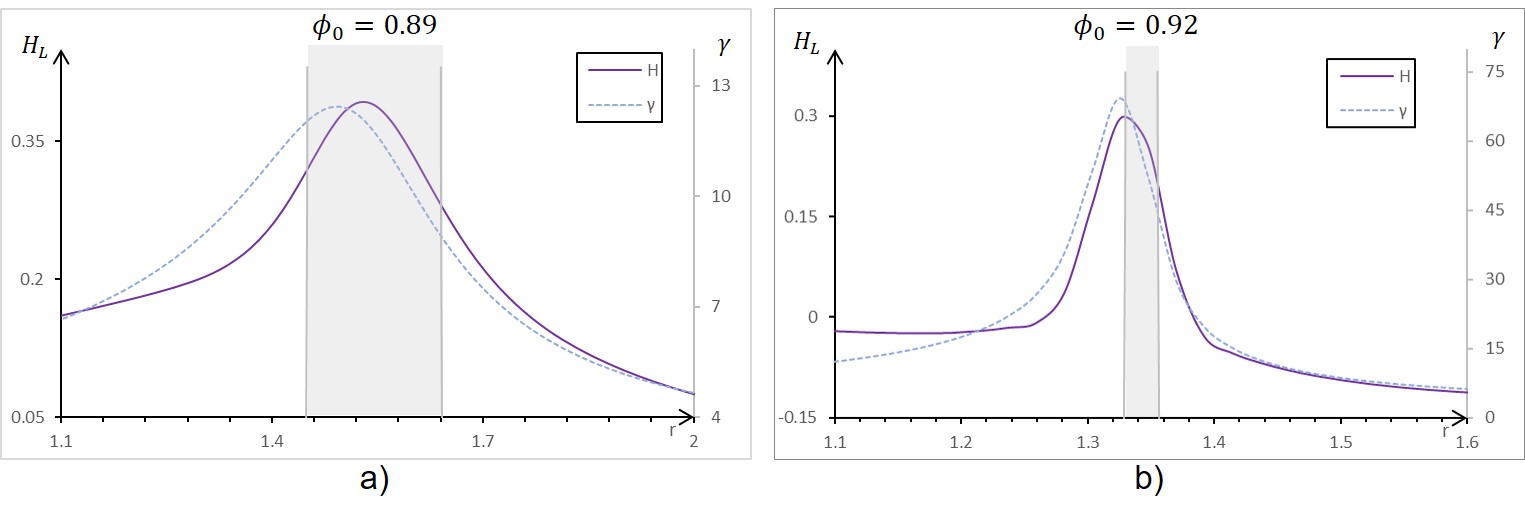}
\caption{\textbf{The longitudinal Hubble rate $H_L$ and $\gamma$
with respect to $r$} across the bubble wall for different initial
conditions. The solid line is the longitudinal Hubble rate (value
shown on the left vertical axis) and the dotted line is $\gamma$
(value shown on the right axis).}
    \label{HL}
\end{figure}

It can be confirmed that in such AdS landscapes, inflation might
occur. According to Refs.\cite{Vilenkin:1994pv,Linde:1994hy}, the
width of the wall is $\delta_0\sim (\phi_2-\phi_1)V_b^{-1/2}$, and
\be \delta_0>{1\over H_0}\simeq \lf({3\over
V_b}\rt)^{1/2}\label{delta0}\ee must be satisfied for the
inflation occurring on the wall. Here, considering the
longitudinal Hubble parameter $H_L = -u^i u^j K_{ij}$ ($u^i$ is
the normal vector at the bubble wall), and $ds^2 = \gamma_{ij} e^i
e^j dr^2 = \gamma^2 dr^2$ ($r$ is the coordinate crossing the
bubble wall and $e^i$ is the direction vector at the wall), we
have \be {\delta_0\simeq \gamma\Delta r }\gtrsim {1\over H_L}, \ee
see Fig.\ref{HL}, which confirms the condition (\ref{delta0}).

\section{Conclusions}

We performed full relativistic simulations for non-perturbative
evolution of spacetime in a simple AdS landscape, and found that
large inhomogeneity of a scalar field inclines to develope into
sphere-like bubbles, and more unexpectedly, into tube-like
structures. It is observed that the spacetime evolution is
completely different in the AdS landscape, the bubble or tube
walls might inflate forever as a quasi-dS space, while the
different regions separated by the walls are in different AdS
vacua and will collapse towards singularity.

Though we take a simple potential with two AdS minima as example,
actually for a spacetime in a landscape with multiple AdS vacua,
its local region always can be described as that with only two
neighbouring AdS minima.
%Although the phenomenology in a multiple-AdS landscape seems to be
%more complex, but locally it can be essentially similar to that in
%our simplified AdS potential.
Thus our results might have effectively captured some of the
essentials of non-perturbative phenomenology of spacetime in AdS
landscape. In this sense, further research on the NR simulations
in a more realistic landscape for the early universe is
interesting and required, which will help to deepen insight into
the origin of our observable Universe.
\\

%According to our simulation results, inflation will inevitably
%occur in AdS landscapes, and the corresponding wall inflation can
%continue forever, which will spawn new inflationary regions and
%thus lead to an eternally inflating multiverse, as pointed out in
%Ref.\cite{Blanco-Pillado:2019tdf}. In such a scenario, the
%inflating region can be surrounded by AdS spacetime, see also
%\cite{Lin:2021ubu}, thus it is possible to encode the information
%of inflation in the dual CFT \cite{Freivogel:2005qh}, see also
%\cite{Levine:2022wos}.

\textbf{Acknowledgment} This work is supported by the NSFC
No.12075246 and by the Fundamental Research Funds for the Central
Universities. We acknowledge the use of GRChombo code for our
simulation and the Tianhe-2 supercomputer for providing computing
resources. We would like to thank Tiago França, Hao-Hao Li and
Hao-Yang Liu for useful discussions.

\bibliography{references}

\begin{thebibliography}{42}
\expandafter\ifx\csname natexlab\endcsname\relax\def\natexlab#1{#1}\fi
\expandafter\ifx\csname bibnamefont\endcsname\relax
  \def\bibnamefont#1{#1}\fi
\expandafter\ifx\csname bibfnamefont\endcsname\relax
  \def\bibfnamefont#1{#1}\fi
\expandafter\ifx\csname citenamefont\endcsname\relax
  \def\citenamefont#1{#1}\fi
\expandafter\ifx\csname url\endcsname\relax
  \def\url#1{\texttt{#1}}\fi
\expandafter\ifx\csname urlprefix\endcsname\relax\def\urlprefix{URL }\fi
\providecommand{\bibinfo}[2]{#2}
\providecommand{\eprint}[2][]{\url{#2}}

\bibitem[{\citenamefont{Guth}(1981)}]{Guth:1980zm}
\bibinfo{author}{\bibfnamefont{A.~H.} \bibnamefont{Guth}}, \bibinfo{journal}{Phys. Rev. D} \textbf{\bibinfo{volume}{23}}, \bibinfo{pages}{347} (\bibinfo{year}{1981}).

\bibitem[{\citenamefont{Linde}(1982)}]{Linde:1981mu}
\bibinfo{author}{\bibfnamefont{A.~D.} \bibnamefont{Linde}}, \bibinfo{journal}{Phys. Lett. B} \textbf{\bibinfo{volume}{108}}, \bibinfo{pages}{389} (\bibinfo{year}{1982}).

\bibitem[{\citenamefont{Albrecht and Steinhardt}(1982)}]{Albrecht:1982wi}
\bibinfo{author}{\bibfnamefont{A.}~\bibnamefont{Albrecht}} \bibnamefont{and} \bibinfo{author}{\bibfnamefont{P.~J.} \bibnamefont{Steinhardt}}, \bibinfo{journal}{Phys. Rev. Lett.} \textbf{\bibinfo{volume}{48}}, \bibinfo{pages}{1220} (\bibinfo{year}{1982}).

\bibitem[{\citenamefont{Starobinsky}(1980)}]{Starobinsky:1980te}
\bibinfo{author}{\bibfnamefont{A.~A.} \bibnamefont{Starobinsky}}, \bibinfo{journal}{Phys. Lett. B} \textbf{\bibinfo{volume}{91}}, \bibinfo{pages}{99} (\bibinfo{year}{1980}).

\bibitem[{\citenamefont{Linde}(1983)}]{Linde:1983gd}
\bibinfo{author}{\bibfnamefont{A.~D.} \bibnamefont{Linde}}, \bibinfo{journal}{Phys. Lett. B} \textbf{\bibinfo{volume}{129}}, \bibinfo{pages}{177} (\bibinfo{year}{1983}).

\bibitem[{\citenamefont{Kachru et~al.}(2003)\citenamefont{Kachru, Kallosh, Linde, and Trivedi}}]{Kachru:2003aw}
\bibinfo{author}{\bibfnamefont{S.}~\bibnamefont{Kachru}}, \bibinfo{author}{\bibfnamefont{R.}~\bibnamefont{Kallosh}}, \bibinfo{author}{\bibfnamefont{A.~D.} \bibnamefont{Linde}}, \bibnamefont{and} \bibinfo{author}{\bibfnamefont{S.~P.} \bibnamefont{Trivedi}}, \bibinfo{journal}{Phys. Rev. D} \textbf{\bibinfo{volume}{68}}, \bibinfo{pages}{046005} (\bibinfo{year}{2003}), \eprint{hep-th/0301240}.

\bibitem[{\citenamefont{Kallosh et~al.}(2019)\citenamefont{Kallosh, Linde, McDonough, and Scalisi}}]{Kallosh:2019axr}
\bibinfo{author}{\bibfnamefont{R.}~\bibnamefont{Kallosh}}, \bibinfo{author}{\bibfnamefont{A.}~\bibnamefont{Linde}}, \bibinfo{author}{\bibfnamefont{E.}~\bibnamefont{McDonough}}, \bibnamefont{and} \bibinfo{author}{\bibfnamefont{M.}~\bibnamefont{Scalisi}}, \bibinfo{journal}{JHEP} \textbf{\bibinfo{volume}{03}}, \bibinfo{pages}{134} (\bibinfo{year}{2019}), \eprint{1901.02022}.

\bibitem[{\citenamefont{Ooguri and Vafa}(2007)}]{Ooguri:2006in}
\bibinfo{author}{\bibfnamefont{H.}~\bibnamefont{Ooguri}} \bibnamefont{and} \bibinfo{author}{\bibfnamefont{C.}~\bibnamefont{Vafa}}, \bibinfo{journal}{Nucl. Phys. B} \textbf{\bibinfo{volume}{766}}, \bibinfo{pages}{21} (\bibinfo{year}{2007}), \eprint{hep-th/0605264}.

\bibitem[{\citenamefont{Obied et~al.}(2018)\citenamefont{Obied, Ooguri, Spodyneiko, and Vafa}}]{Obied:2018sgi}
\bibinfo{author}{\bibfnamefont{G.}~\bibnamefont{Obied}}, \bibinfo{author}{\bibfnamefont{H.}~\bibnamefont{Ooguri}}, \bibinfo{author}{\bibfnamefont{L.}~\bibnamefont{Spodyneiko}}, \bibnamefont{and} \bibinfo{author}{\bibfnamefont{C.}~\bibnamefont{Vafa}} (\bibinfo{year}{2018}), \eprint{1806.08362}.

\bibitem[{\citenamefont{Brennan et~al.}(2017)\citenamefont{Brennan, Carta, and Vafa}}]{Brennan:2017rbf}
\bibinfo{author}{\bibfnamefont{T.~D.} \bibnamefont{Brennan}}, \bibinfo{author}{\bibfnamefont{F.}~\bibnamefont{Carta}}, \bibnamefont{and} \bibinfo{author}{\bibfnamefont{C.}~\bibnamefont{Vafa}}, \bibinfo{journal}{PoS} \textbf{\bibinfo{volume}{TASI2017}}, \bibinfo{pages}{015} (\bibinfo{year}{2017}), \eprint{1711.00864}.

\bibitem[{\citenamefont{Palti}(2022)}]{Palti:2022edh}
\bibinfo{author}{\bibfnamefont{E.}~\bibnamefont{Palti}}, \bibinfo{journal}{Contemp. Phys.} \textbf{\bibinfo{volume}{62}}, \bibinfo{pages}{165} (\bibinfo{year}{2022}).

\bibitem[{\citenamefont{Chen et~al.}(2021)\citenamefont{Chen, Gorbenko, and Maldacena}}]{Chen:2020tes}
\bibinfo{author}{\bibfnamefont{Y.}~\bibnamefont{Chen}}, \bibinfo{author}{\bibfnamefont{V.}~\bibnamefont{Gorbenko}}, \bibnamefont{and} \bibinfo{author}{\bibfnamefont{J.}~\bibnamefont{Maldacena}}, \bibinfo{journal}{JHEP} \textbf{\bibinfo{volume}{02}}, \bibinfo{pages}{009} (\bibinfo{year}{2021}), \eprint{2007.16091}.

\bibitem[{\citenamefont{Hartman et~al.}(2020)\citenamefont{Hartman, Jiang, and Shaghoulian}}]{Hartman:2020khs}
\bibinfo{author}{\bibfnamefont{T.}~\bibnamefont{Hartman}}, \bibinfo{author}{\bibfnamefont{Y.}~\bibnamefont{Jiang}}, \bibnamefont{and} \bibinfo{author}{\bibfnamefont{E.}~\bibnamefont{Shaghoulian}}, \bibinfo{journal}{JHEP} \textbf{\bibinfo{volume}{11}}, \bibinfo{pages}{111} (\bibinfo{year}{2020}), \eprint{2008.01022}.

\bibitem[{\citenamefont{Levine and Shaghoulian}(2022)}]{Levine:2022wos}
\bibinfo{author}{\bibfnamefont{A.}~\bibnamefont{Levine}} \bibnamefont{and} \bibinfo{author}{\bibfnamefont{E.}~\bibnamefont{Shaghoulian}} (\bibinfo{year}{2022}), \eprint{2204.08503}.

\bibitem[{\citenamefont{Cooper et~al.}(2019)\citenamefont{Cooper, Rozali, Swingle, Van~Raamsdonk, Waddell, and Wakeham}}]{Cooper:2018cmb}
\bibinfo{author}{\bibfnamefont{S.}~\bibnamefont{Cooper}}, \bibinfo{author}{\bibfnamefont{M.}~\bibnamefont{Rozali}}, \bibinfo{author}{\bibfnamefont{B.}~\bibnamefont{Swingle}}, \bibinfo{author}{\bibfnamefont{M.}~\bibnamefont{Van~Raamsdonk}}, \bibinfo{author}{\bibfnamefont{C.}~\bibnamefont{Waddell}}, \bibnamefont{and} \bibinfo{author}{\bibfnamefont{D.}~\bibnamefont{Wakeham}}, \bibinfo{journal}{JHEP} \textbf{\bibinfo{volume}{07}}, \bibinfo{pages}{065} (\bibinfo{year}{2019}), \eprint{1810.10601}.

\bibitem[{\citenamefont{Antonini et~al.}(2022{\natexlab{a}})\citenamefont{Antonini, Simidzija, Swingle, and Van~Raamsdonk}}]{Antonini:2022blk}
\bibinfo{author}{\bibfnamefont{S.}~\bibnamefont{Antonini}}, \bibinfo{author}{\bibfnamefont{P.}~\bibnamefont{Simidzija}}, \bibinfo{author}{\bibfnamefont{B.}~\bibnamefont{Swingle}}, \bibnamefont{and} \bibinfo{author}{\bibfnamefont{M.}~\bibnamefont{Van~Raamsdonk}} (\bibinfo{year}{2022}{\natexlab{a}}), \eprint{2203.11220}.

\bibitem[{\citenamefont{Antonini et~al.}(2022{\natexlab{b}})\citenamefont{Antonini, Simidzija, Swingle, and Van~Raamsdonk}}]{Antonini:2022xzo}
\bibinfo{author}{\bibfnamefont{S.}~\bibnamefont{Antonini}}, \bibinfo{author}{\bibfnamefont{P.}~\bibnamefont{Simidzija}}, \bibinfo{author}{\bibfnamefont{B.}~\bibnamefont{Swingle}}, \bibnamefont{and} \bibinfo{author}{\bibfnamefont{M.}~\bibnamefont{Van~Raamsdonk}} (\bibinfo{year}{2022}{\natexlab{b}}), \eprint{2206.14821}.

\bibitem[{\citenamefont{Ye and Piao}(2020{\natexlab{a}})}]{Ye:2020btb}
\bibinfo{author}{\bibfnamefont{G.}~\bibnamefont{Ye}} \bibnamefont{and} \bibinfo{author}{\bibfnamefont{Y.-S.} \bibnamefont{Piao}}, \bibinfo{journal}{Phys. Rev. D} \textbf{\bibinfo{volume}{101}}, \bibinfo{pages}{083507} (\bibinfo{year}{2020}{\natexlab{a}}), \eprint{2001.02451}.

\bibitem[{\citenamefont{Ye and Piao}(2020{\natexlab{b}})}]{Ye:2020oix}
\bibinfo{author}{\bibfnamefont{G.}~\bibnamefont{Ye}} \bibnamefont{and} \bibinfo{author}{\bibfnamefont{Y.-S.} \bibnamefont{Piao}}, \bibinfo{journal}{Phys. Rev. D} \textbf{\bibinfo{volume}{102}}, \bibinfo{pages}{083523} (\bibinfo{year}{2020}{\natexlab{b}}), \eprint{2008.10832}.

\bibitem[{\citenamefont{Jiang and Piao}(2021)}]{Jiang:2021bab}
\bibinfo{author}{\bibfnamefont{J.-Q.} \bibnamefont{Jiang}} \bibnamefont{and} \bibinfo{author}{\bibfnamefont{Y.-S.} \bibnamefont{Piao}}, \bibinfo{journal}{Phys. Rev. D} \textbf{\bibinfo{volume}{104}}, \bibinfo{pages}{103524} (\bibinfo{year}{2021}), \eprint{2107.07128}.

\bibitem[{\citenamefont{Ye et~al.}(2021)\citenamefont{Ye, Zhang, and Piao}}]{Ye:2021iwa}
\bibinfo{author}{\bibfnamefont{G.}~\bibnamefont{Ye}}, \bibinfo{author}{\bibfnamefont{J.}~\bibnamefont{Zhang}}, \bibnamefont{and} \bibinfo{author}{\bibfnamefont{Y.-S.} \bibnamefont{Piao}} (\bibinfo{year}{2021}), \eprint{2107.13391}.

\bibitem[{\citenamefont{Feldbrugge et~al.}(2017)\citenamefont{Feldbrugge, Lehners, and Turok}}]{Feldbrugge:2017fcc}
\bibinfo{author}{\bibfnamefont{J.}~\bibnamefont{Feldbrugge}}, \bibinfo{author}{\bibfnamefont{J.-L.} \bibnamefont{Lehners}}, \bibnamefont{and} \bibinfo{author}{\bibfnamefont{N.}~\bibnamefont{Turok}}, \bibinfo{journal}{Phys. Rev. Lett.} \textbf{\bibinfo{volume}{119}}, \bibinfo{pages}{171301} (\bibinfo{year}{2017}), \eprint{1705.00192}.

\bibitem[{\citenamefont{Borde et~al.}(2003)\citenamefont{Borde, Guth, and Vilenkin}}]{Borde:2001nh}
\bibinfo{author}{\bibfnamefont{A.}~\bibnamefont{Borde}}, \bibinfo{author}{\bibfnamefont{A.~H.} \bibnamefont{Guth}}, \bibnamefont{and} \bibinfo{author}{\bibfnamefont{A.}~\bibnamefont{Vilenkin}}, \bibinfo{journal}{Phys. Rev. Lett.} \textbf{\bibinfo{volume}{90}}, \bibinfo{pages}{151301} (\bibinfo{year}{2003}), \eprint{gr-qc/0110012}.

\bibitem[{\citenamefont{Pretorius}(2005)}]{Pretorius:2005gq}
\bibinfo{author}{\bibfnamefont{F.}~\bibnamefont{Pretorius}}, \bibinfo{journal}{Phys. Rev. Lett.} \textbf{\bibinfo{volume}{95}}, \bibinfo{pages}{121101} (\bibinfo{year}{2005}), \eprint{gr-qc/0507014}.

\bibitem[{\citenamefont{Campanelli et~al.}(2006)\citenamefont{Campanelli, Lousto, Marronetti, and Zlochower}}]{Campanelli:2005dd}
\bibinfo{author}{\bibfnamefont{M.}~\bibnamefont{Campanelli}}, \bibinfo{author}{\bibfnamefont{C.~O.} \bibnamefont{Lousto}}, \bibinfo{author}{\bibfnamefont{P.}~\bibnamefont{Marronetti}}, \bibnamefont{and} \bibinfo{author}{\bibfnamefont{Y.}~\bibnamefont{Zlochower}}, \bibinfo{journal}{Phys. Rev. Lett.} \textbf{\bibinfo{volume}{96}}, \bibinfo{pages}{111101} (\bibinfo{year}{2006}), \eprint{gr-qc/0511048}.

\bibitem[{\citenamefont{Baker et~al.}(2006)\citenamefont{Baker, Centrella, Choi, Koppitz, and van Meter}}]{Baker:2005vv}
\bibinfo{author}{\bibfnamefont{J.~G.} \bibnamefont{Baker}}, \bibinfo{author}{\bibfnamefont{J.}~\bibnamefont{Centrella}}, \bibinfo{author}{\bibfnamefont{D.-I.} \bibnamefont{Choi}}, \bibinfo{author}{\bibfnamefont{M.}~\bibnamefont{Koppitz}}, \bibnamefont{and} \bibinfo{author}{\bibfnamefont{J.}~\bibnamefont{van Meter}}, \bibinfo{journal}{Phys. Rev. Lett.} \textbf{\bibinfo{volume}{96}}, \bibinfo{pages}{111102} (\bibinfo{year}{2006}), \eprint{gr-qc/0511103}.

\bibitem[{\citenamefont{Giblin et~al.}(2016)\citenamefont{Giblin, Mertens, and Starkman}}]{Giblin:2015vwq}
\bibinfo{author}{\bibfnamefont{J.~T.} \bibnamefont{Giblin}}, \bibinfo{author}{\bibfnamefont{J.~B.} \bibnamefont{Mertens}}, \bibnamefont{and} \bibinfo{author}{\bibfnamefont{G.~D.} \bibnamefont{Starkman}}, \bibinfo{journal}{Phys. Rev. Lett.} \textbf{\bibinfo{volume}{116}}, \bibinfo{pages}{251301} (\bibinfo{year}{2016}), \eprint{1511.01105}.

\bibitem[{\citenamefont{Bentivegna and Bruni}(2016)}]{Bentivegna:2015flc}
\bibinfo{author}{\bibfnamefont{E.}~\bibnamefont{Bentivegna}} \bibnamefont{and} \bibinfo{author}{\bibfnamefont{M.}~\bibnamefont{Bruni}}, \bibinfo{journal}{Phys. Rev. Lett.} \textbf{\bibinfo{volume}{116}}, \bibinfo{pages}{251302} (\bibinfo{year}{2016}), \eprint{1511.05124}.

\bibitem[{\citenamefont{Musoke et~al.}(2020)\citenamefont{Musoke, Hotchkiss, and Easther}}]{Musoke:2019ima}
\bibinfo{author}{\bibfnamefont{N.}~\bibnamefont{Musoke}}, \bibinfo{author}{\bibfnamefont{S.}~\bibnamefont{Hotchkiss}}, \bibnamefont{and} \bibinfo{author}{\bibfnamefont{R.}~\bibnamefont{Easther}}, \bibinfo{journal}{Phys. Rev. Lett.} \textbf{\bibinfo{volume}{124}}, \bibinfo{pages}{061301} (\bibinfo{year}{2020}), \eprint{1909.11678}.

\bibitem[{\citenamefont{Macpherson et~al.}(2017)\citenamefont{Macpherson, Lasky, and Price}}]{Macpherson:2016ict}
\bibinfo{author}{\bibfnamefont{H.~J.} \bibnamefont{Macpherson}}, \bibinfo{author}{\bibfnamefont{P.~D.} \bibnamefont{Lasky}}, \bibnamefont{and} \bibinfo{author}{\bibfnamefont{D.~J.} \bibnamefont{Price}}, \bibinfo{journal}{Phys. Rev. D} \textbf{\bibinfo{volume}{95}}, \bibinfo{pages}{064028} (\bibinfo{year}{2017}), \eprint{1611.05447}.

\bibitem[{\citenamefont{Macpherson et~al.}(2019)\citenamefont{Macpherson, Price, and Lasky}}]{Macpherson:2018btl}
\bibinfo{author}{\bibfnamefont{H.~J.} \bibnamefont{Macpherson}}, \bibinfo{author}{\bibfnamefont{D.~J.} \bibnamefont{Price}}, \bibnamefont{and} \bibinfo{author}{\bibfnamefont{P.~D.} \bibnamefont{Lasky}}, \bibinfo{journal}{Phys. Rev. D} \textbf{\bibinfo{volume}{99}}, \bibinfo{pages}{063522} (\bibinfo{year}{2019}), \eprint{1807.01711}.

\bibitem[{\citenamefont{Abbott and Coleman}(1985)}]{Abbott:1985kr}
\bibinfo{author}{\bibfnamefont{L.~F.} \bibnamefont{Abbott}} \bibnamefont{and} \bibinfo{author}{\bibfnamefont{S.~R.} \bibnamefont{Coleman}}, \bibinfo{journal}{Nucl. Phys. B} \textbf{\bibinfo{volume}{259}}, \bibinfo{pages}{170} (\bibinfo{year}{1985}).

\bibitem[{\citenamefont{Bizon and Rostworowski}(2011)}]{Bizon:2011gg}
\bibinfo{author}{\bibfnamefont{P.}~\bibnamefont{Bizon}} \bibnamefont{and} \bibinfo{author}{\bibfnamefont{A.}~\bibnamefont{Rostworowski}}, \bibinfo{journal}{Phys. Rev. Lett.} \textbf{\bibinfo{volume}{107}}, \bibinfo{pages}{031102} (\bibinfo{year}{2011}), \eprint{1104.3702}.

\bibitem[{\citenamefont{Blanco-Pillado et~al.}(2020)\citenamefont{Blanco-Pillado, Deng, and Vilenkin}}]{Blanco-Pillado:2019tdf}
\bibinfo{author}{\bibfnamefont{J.~J.} \bibnamefont{Blanco-Pillado}}, \bibinfo{author}{\bibfnamefont{H.}~\bibnamefont{Deng}}, \bibnamefont{and} \bibinfo{author}{\bibfnamefont{A.}~\bibnamefont{Vilenkin}}, \bibinfo{journal}{JCAP} \textbf{\bibinfo{volume}{05}}, \bibinfo{pages}{014} (\bibinfo{year}{2020}), \eprint{1909.00068}.

\bibitem[{\citenamefont{East et~al.}(2016)\citenamefont{East, Kleban, Linde, and Senatore}}]{East:2015ggf}
\bibinfo{author}{\bibfnamefont{W.~E.} \bibnamefont{East}}, \bibinfo{author}{\bibfnamefont{M.}~\bibnamefont{Kleban}}, \bibinfo{author}{\bibfnamefont{A.}~\bibnamefont{Linde}}, \bibnamefont{and} \bibinfo{author}{\bibfnamefont{L.}~\bibnamefont{Senatore}}, \bibinfo{journal}{JCAP} \textbf{\bibinfo{volume}{09}}, \bibinfo{pages}{010} (\bibinfo{year}{2016}), \eprint{1511.05143}.

\bibitem[{\citenamefont{Clough et~al.}(2017)\citenamefont{Clough, Lim, DiNunno, Fischler, Flauger, and Paban}}]{Clough:2016ymm}
\bibinfo{author}{\bibfnamefont{K.}~\bibnamefont{Clough}}, \bibinfo{author}{\bibfnamefont{E.~A.} \bibnamefont{Lim}}, \bibinfo{author}{\bibfnamefont{B.~S.} \bibnamefont{DiNunno}}, \bibinfo{author}{\bibfnamefont{W.}~\bibnamefont{Fischler}}, \bibinfo{author}{\bibfnamefont{R.}~\bibnamefont{Flauger}}, \bibnamefont{and} \bibinfo{author}{\bibfnamefont{S.}~\bibnamefont{Paban}}, \bibinfo{journal}{JCAP} \textbf{\bibinfo{volume}{09}}, \bibinfo{pages}{025} (\bibinfo{year}{2017}), \eprint{1608.04408}.

\bibitem[{\citenamefont{Lin and Piao}(2022)}]{Lin:2021ubu}
\bibinfo{author}{\bibfnamefont{P.-X.} \bibnamefont{Lin}} \bibnamefont{and} \bibinfo{author}{\bibfnamefont{Y.-S.} \bibnamefont{Piao}}, \bibinfo{journal}{Phys. Rev. D} \textbf{\bibinfo{volume}{105}}, \bibinfo{pages}{063534} (\bibinfo{year}{2022}), \eprint{2111.09174}.

\bibitem[{\citenamefont{Baumgarte and Shapiro}(1998)}]{Baumgarte:1998te}
\bibinfo{author}{\bibfnamefont{T.~W.} \bibnamefont{Baumgarte}} \bibnamefont{and} \bibinfo{author}{\bibfnamefont{S.~L.} \bibnamefont{Shapiro}}, \bibinfo{journal}{Phys. Rev. D} \textbf{\bibinfo{volume}{59}}, \bibinfo{pages}{024007} (\bibinfo{year}{1998}), \eprint{gr-qc/9810065}.

\bibitem[{\citenamefont{Shibata and Nakamura}(1995)}]{Shibata:1995we}
\bibinfo{author}{\bibfnamefont{M.}~\bibnamefont{Shibata}} \bibnamefont{and} \bibinfo{author}{\bibfnamefont{T.}~\bibnamefont{Nakamura}}, \bibinfo{journal}{Phys. Rev. D} \textbf{\bibinfo{volume}{52}}, \bibinfo{pages}{5428} (\bibinfo{year}{1995}).

\bibitem[{\citenamefont{Clough et~al.}(2015)\citenamefont{Clough, Figueras, Finkel, Kunesch, Lim, and Tunyasuvunakool}}]{Clough:2015sqa}
\bibinfo{author}{\bibfnamefont{K.}~\bibnamefont{Clough}}, \bibinfo{author}{\bibfnamefont{P.}~\bibnamefont{Figueras}}, \bibinfo{author}{\bibfnamefont{H.}~\bibnamefont{Finkel}}, \bibinfo{author}{\bibfnamefont{M.}~\bibnamefont{Kunesch}}, \bibinfo{author}{\bibfnamefont{E.~A.} \bibnamefont{Lim}}, \bibnamefont{and} \bibinfo{author}{\bibfnamefont{S.}~\bibnamefont{Tunyasuvunakool}}, \bibinfo{journal}{Class. Quant. Grav.} \textbf{\bibinfo{volume}{32}}, \bibinfo{pages}{245011} (\bibinfo{year}{2015}), \eprint{1503.03436}.

\bibitem[{\citenamefont{Vilenkin}(1994)}]{Vilenkin:1994pv}
\bibinfo{author}{\bibfnamefont{A.}~\bibnamefont{Vilenkin}}, \bibinfo{journal}{Phys. Rev. Lett.} \textbf{\bibinfo{volume}{72}}, \bibinfo{pages}{3137} (\bibinfo{year}{1994}), \eprint{hep-th/9402085}.

\bibitem[{\citenamefont{Linde}(1994)}]{Linde:1994hy}
\bibinfo{author}{\bibfnamefont{A.~D.} \bibnamefont{Linde}}, \bibinfo{journal}{Phys. Lett. B} \textbf{\bibinfo{volume}{327}}, \bibinfo{pages}{208} (\bibinfo{year}{1994}), \eprint{astro-ph/9402031}.

\end{thebibliography}

%\appendix

\end{document}